\documentclass[aps,prb,12pt,superscriptaddress,letter]{revtex4}
\usepackage{amsfonts}
\usepackage{amsmath}
\usepackage{amssymb}
\usepackage{graphicx}

\begin{document}

\title{Quantum Phase Transitions of Topological Insulators}
\author{Lan-Feng Liu }
\affiliation{Department of Physics, Beijing Normal University, Beijing 100875, China}
\author{Su-Peng Kou}
\thanks{Corresponding author}
\email{spkou@bnu.edu.cn}
\affiliation{Department of Physics, Beijing Normal University, Beijing 100875, China}
\pacs{72.10.Bg, 71.70.Ej,
72.25.-b}
\date{\today }

\begin{abstract}
In this paper, starting from a lattice model of topological insulators, we
study the quantum phase transitions among different quantum states,
including quantum spin Hall state, quantum anomalous Hall state and normal
band insulator state by calculating their topological properties (edge
states, quantized spin Hall conductivities and the number of zero mode on a $%
\pi $-flux). We find that there exist universal features for the topological
quantum phase transitions (TQPTs) in different cases : the emergence of
nodal fermions at high symmetry points, the non-analytic third derivative of
ground state energy and the jumps of the topological "order parameters". In
particular, the relation between TQPTs and symmetries of the systems are
explored : different TQPTs are protected by different (global) symmetries
and then described by different topological "order parameters".
\end{abstract}

\pacs{73.50.Mx, 78.66.-w, 81.05.Uw}
\maketitle

\section{\textbf{Introduction}}

As the first topologically ordered phase, the Quantum Hall Effect (QHE) is a
remarkable achievement in condensed matter physics\cite{2,qhe}. In QHE
state, at low temperatures and in strong magnetic fields, quantized Hall
conductance can be observed due to the Landau levels formed by states of a
two-dimensional electron gas. Just one year later, people discovered the
so-called anomalous Hall effect where additional Hall current in the
ferromagnetic material was observed\cite{3,fang}. Then, people find that the
transverse transport also exists for different spin species, which is called
as spin Hall effect\cite{zhang1}. Recently, the topological insulators
including the quantized spin Hall (QSH) effect\cite{kane,zhang,berg,kon} and
the quantized anomalous Hall (QAH) effect\cite{haldane,liu} become hot
issues.

On the one hand, to describe the quantized anomalous Hall state, a class of
topological insulators with time-reversal symmetry breaking, the Chern
number, so called TKNN integer, is introduced as integrals over the
Brillouin zone (BZ) of the Berry field strength \cite{thou}
\begin{equation}
Q=-\frac{1}{2\pi }\int_{BZ}(\mathbf{\nabla }_{k}\times \mathbf{A}%
)_{z}\,dk_{x}dk_{y}.
\end{equation}%
Here $A_{j}(k)=-i\langle u_{j}|\nabla _{k}|u_{j}\rangle $ is a Berry
connection for single-electron Hamiltonian with the periodic part $u(k)$ of
a Bloch state $\psi _{\mathbf{k}}=u(k)e^{i\mathbf{k}\cdot \mathbf{r}}$.

On the other hand, for quantized spin Hall state the spin Chern number $%
Q_{s}=Q_{\uparrow }-Q_{\downarrow }$ is proposed in Ref.\cite{sheng} to be
the topological invariant to characterize the topological insulators.
However, due to spin mixing term (the Rashba term), the spin Chern number $%
Q_{s}$ is not well defined. Thus for a class of topological insulators with
time-reversal symmetry (T-symmetry), due to Kramers degeneracy, Kane and
Mele proposed a $Z_{2}$ topological invariant\cite%
{kane,Roy06,Moore06,Fu06,Essin07}
\begin{equation}
(-1)^{\Delta }=\prod_{i=1}^{4}\prod_{m=1}^{N}\xi _{2m}(\vec{k}_{i}),
\end{equation}%
where $\vec{k}_{i}$ are the four high-symmetry points satisfying $\vec{k}%
=(0,0)$, $(0,\pi )$, $(\pi ,0)$, $(\pi ,\pi )$, $\xi _{2m}(\vec{k}_{i})$ is
the parity eigenvalue at each of these points, and $N$ is the number of
Kramers pairs below Fermi surface. Such $Z_{2}$ topological invariant can be
also defined in terms of the spin Chern number
\begin{equation}
\Delta =\frac{1}{2\pi }\int_{EBZ}d^{2}k\,(\mathbf{\nabla }_{k}\times \mathbf{%
A})_{z}\text{\textrm{mod}}2.  \label{Dk}
\end{equation}%
Here the integral of the Berry field strength is defined on half of the
Brillouin zone. Physically, $\Delta $ is identical with the number of pairs
of helical edge modes.

Thus, an interesting issue is the nature of the quantum phase transitions
between different types of topological insulators. It is well known that in
Landau theory different orders are classified by \emph{symmetries}. The
phase transitions accompanied with (global) symmetry breaking are always
described by order parameters. However, Landau's theory fails to describe
the topological quantum phase transition (TQPT). Such type of quantum phase
transition cannot be classified by symmetries\cite{wen}. Instead, they may
be characterized by some topological "order parameters", such as the Chern
number or $Z_{2}$ topological invariant\cite{lifshiz,volovik,qi1}.

In this paper, we study the TQPTs between topological insulators and show
their physical properties. We find that there are universal features of the
TQPTs for different cases : the existence of nodal fermions at high symmetry
points, the non-analytic third derivative of ground state energy and the
jumps of the topological "order parameters". In particular, we find the
\emph{symmetry protected} nature of the TQPTs : different TQPTs are
protected by different (global) symmetries and then described by different
topological "order parameters".

The remainder of the paper is organized as follows. In Sec.II, the
two-dimensional lattice models of topological insulators are given. In
Sec.III, we focus on the TQPTs between different quantum states with $S^{z}$%
-conservation (without time reversal symmetry). In this section, the global
phase diagram and the critical behavior are obtained. In addition, the
topological properties of different quantum states are calculated, including
the edge states, the quantized Hall conductivities and the induced quantum
numbers on a $\pi $-flux. In Sec.IV, we will study the TQPTs with time
reversal symmetry (without $S^{z}$-conservation). In Sec.V, We will study
the TQPTs without time reversal symmetry and $S^{z}$-conservation. Finally,
the conclusions are given in Sec.VI.

\section{The lattice models of topological insulators}

As a starting point, we consider a lattice Hamiltonian for two-flavor
spin-1/2 fermions\cite{com}:

\begin{equation}
H_{1}=H_{0}+H_{D},
\end{equation}%
where%
\begin{eqnarray}
H_{0} &=&\underset{i,\sigma }{\sum }\Psi _{i,\sigma }^{\dagger }\left(
\begin{array}{cc}
-\mu /2 & 0 \\
0 & \mu /2%
\end{array}%
\right) \Psi _{i,\sigma }+\underset{i,\sigma }{\sum }\tau _{i,i+x}^{z}\Psi
_{i+x,\sigma }^{\dagger }\left(
\begin{array}{cc}
-t & -1/2 \\
1/2 & t%
\end{array}%
\right) \Psi _{i,\sigma }  \notag \\
&&\underset{i,\sigma }{+\sum }\tau _{i,i+y}^{z}\Psi _{i+y,\sigma }^{\dagger
}\left(
\begin{array}{cc}
-t & i\mathrm{sgn}(\sigma )/2 \\
i\mathrm{sgn}(\sigma )/2 & t%
\end{array}%
\right) \Psi _{i,\sigma }+H.c., \\
H_{D} &=&\underset{i,\sigma }{D\sum }\Psi _{i,\sigma }^{\dagger }\left(
\begin{array}{cc}
-\mathrm{sgn}(\sigma ) & 0 \\
0 & \mathrm{sgn}(\sigma )%
\end{array}%
\right) \Psi _{i,\sigma }.
\end{eqnarray}%
Here $\Psi _{i,\sigma }^{\dagger }=(\psi _{i,A\sigma }^{\dagger },\psi
_{i,B\sigma }^{\dagger })$ is a two-component particle operator where $%
\sigma \equiv \uparrow ,\downarrow $ is the spin index, $A$, $B$ denote the
flavor indices and $i$\ labels the site on the square lattice. $t$ is the
real hopping parameters. $\mu $ is the orbital splitting energy. $D$ is an
"effective" magnetic field which was first introduced in Ref.\cite{liu}
phenomenologically which may be induced by local magnetic moments. The bond
variables $\tau _{i,j}^{z}$ is set to $+1$ everywhere at the beginning. For
the case of $D>0$, we set the $\mathrm{sgn}(\sigma )D$ is positive for
spin-up while the $\mathrm{sgn}(\sigma )D$ is negative for spin down.
Without $D$ terms, the model in Eq.[3] is really a lattice realization of
the Kane-Mele model proposed in Ref.\cite{ran,ran1}, of which the ground
state is a topological insulator for $t>1/4$.

Using the Fourier transform $\Psi _{i}=\frac{1}{\sqrt{N}}\underset{k}{\sum }%
e^{-ik\cdot R_{i}}\Psi _{k},$ we can transform the four field operators on
each site into momentum space $\Psi _{k}^{\dagger }=(\Psi _{k,A\uparrow
}^{\dagger },\Psi _{k,B\uparrow }^{\dagger },\Psi _{k,A\downarrow }^{\dagger
},\Psi _{k,B\downarrow }^{\dagger }).$ Then we get
\begin{equation}
H_{k}=\sum_{k}\Psi _{k}^{\dagger }\left(
\begin{array}{cc}
h_{+(k)} & 0 \\
0 & h_{-(k)}%
\end{array}%
\right) \Psi _{k},
\end{equation}%
where, for the spin up part
\begin{widetext}
\begin{equation*}
h_{+(k)}=\left(
\begin{array}{cc}
-\frac{\mu }{2}-2t(\cos k_{x}+\cos k_{y})-D & \sin k_{y}+i\sin
k_{x} \\
 \sin k_{y}-i\sin k_{x} & \frac{\mu }{2}+2t(\cos k_{x}+\cos k_{y})+D%
\end{array}%
\right) .
\end{equation*}
\end{widetext}
and for the spin down part
\begin{widetext}
\begin{equation*}
h_{-(k)}=\left(
\begin{array}{cc}
-\frac{\mu }{2}-2t(\cos k_{x}+\cos k_{y})+D & -(\sin k_{y}-i\sin
k_{x}) \\
- (\sin k_{y}+i\sin k_{x}) & \frac{\mu }{2}+2t(\cos k_{x}+\cos
k_{y})-D%
\end{array}%
\right) .
\end{equation*}
\end{widetext}
The energy spectrum is then given by diagonalizing the Hamiltonian%
\begin{equation}
E_{k}=\pm \sqrt{(\sin ^{2}k_{y}+\sin ^{2}k_{x})+(\frac{\mu }{2}+2t(\cos
k_{x}+\cos k_{y})\pm D)^{2}}.
\end{equation}%
It is obvious that the term of $\mathrm{sgn}(\sigma )D$ breaks the
T-symmetry,
\begin{equation}
H_{1}(k)\neq \Theta ^{-1}H_{1}^{T}(-k)\Theta
\end{equation}%
where $\Theta =i\sigma _{y}K$ ($\sigma _{x,y,z}$ are Pauli matrices and $K$
stands for complex conjugation). However, the model has an additional spin
rotation symmetry around $z$-direction as
\begin{equation}
H_{1}=e^{-i\theta \sigma _{z}}H_{1}e^{i\theta \sigma _{z}}.
\end{equation}

In addition, by adding the Rashba term to $H_{0}$, we get another lattice
model $H_{2}=H_{0}+H_{R}$ where
\begin{equation}
H_{R}=\sum_{i}R(\Psi _{i,\uparrow }^{\dagger }\Psi _{i+x,\downarrow }-\Psi
_{i,\downarrow }^{\dagger }\Psi _{i+x,\uparrow })+\sum_{i}iR(\Psi
_{i,\uparrow }^{\dagger }\Psi _{i+y,\downarrow }-\Psi _{i\downarrow
,}^{\dagger }\Psi _{i+y,\uparrow })+H.c..
\end{equation}%
By using the Fourier transform we get
\begin{equation}
H_{R}=\sum_{k}\Psi _{k}^{\dagger }\left(
\begin{array}{cc}
0 & R(\sin k_{y}+i\sin k_{x})\tau ^{0} \\
R(\sin k_{y}-i\sin k_{x})\tau ^{0} & 0%
\end{array}%
\right) \Psi _{k},
\end{equation}%
where, $\tau ^{0}$ is the unit matrix. Now there is T-symmetry%
\begin{equation}
H_{2}(k)=\Theta ^{-1}H_{2}^{T}(-k)\Theta .
\end{equation}%
However $S_{z}$ is not a good quantum number due to
\begin{equation}
H_{2}\neq e^{-i\theta \sigma _{z}}H_{2}e^{i\theta \sigma _{z}}.
\end{equation}

Furthermore, one may consider a lattice model $H_{3}$ with both $H_{D}$ and $%
H_{R}$ as $H_{3}=H_{0}+H_{D}+H_{R}$. Now the Hamiltonian $H_{3}$ has neither
T-symmetry nor $S_{z}$-conservation due to
\begin{equation}
H_{3}(k)\neq \Theta ^{-1}H_{3}^{T}(-k)\Theta
\end{equation}%
and%
\begin{equation}
H_{3}\neq e^{-i\theta \sigma _{z}}H_{2}e^{i\theta \sigma _{z}}.
\end{equation}

In the following parts we will study TQPTs by keeping different symmetries
based on different lattice models $H_{1,2,3}$.

\section{TQPTs with $S^{z}$-conservation : $D\neq 0,$ $R=0$}

\subsection{Global phase diagram}

In this section, to learn the TQPT of topological insulators with $S^{z}$%
-conservation ($D\neq 0,$ $R=0$)$,\ $we focus on $H_{1}=H_{0}+H_{D}.$
\begin{figure}[tbph]
\includegraphics[width = 10.0cm]{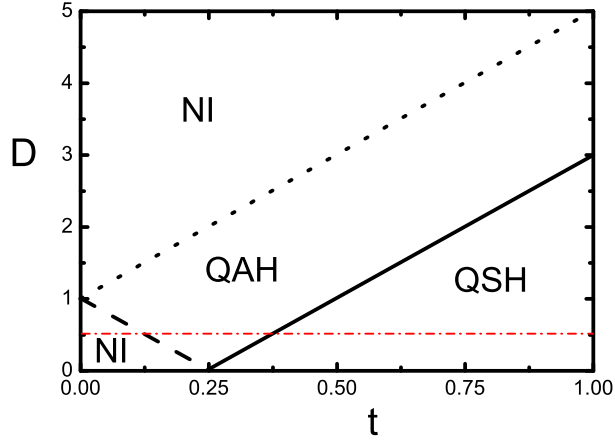}
\caption{The global quantum phase transition diagram. The solid, dashed and
dotted lines are $-\protect\mu /2+4t=D,$ $\protect\mu /2-4t=D,$ $\protect\mu %
/2+4t=D$ respectively. these lines consist of the boundary of the quantum
anomalous Hall state. The solid line is the critical line separating the QAH
and QSH. The dashed and dotted lines are the critical lines separating the
QAH and NI. The dash dot line intersects the solid and dotted lines at the
points $(t,$ $D)=(0.125,$ $0.5)$ and $(t,$ $D)=(0.375,$ $0.5).$ The
parameter $\protect\mu $ is set to be 2.}
\end{figure}

By calculating the phase boundary with zero fermion energy, we obtain the
phase boundary,
\begin{equation}
1\pm 4t=\pm D.
\end{equation}%
In the global phase diagram which was shown in Fig.1, there exist three
phases : QAH state, QSH State and NI state. Taking $D=0.5$ as an example
(the horizontal dash dot line in Fig.1), one can see that the QSH state
occurs in the region of $t>0.375$. Although the T-symmetry is broken in this
region, the topological properties of the QSH state is preserved (the edge
states, the quantized Hall conductivities and the induced quantum numbers on
a $\pi $-flux). With the decrease of $t$, the system turns into a QAH state
in the region of $0.375>t>0.125.$ In this region, one may find that the
topological properties are different from that in the QSH state. In the
region of $0.125>t>0$, the ground state turns into a normal band insulator
with trivial topological properties. In Fig.1, the solid, dashed and dotted
lines are given by $-1+4t=D,$ $1-4t=D$ and $1+4t=D$ which denote the phase
boundaries obtained above.

\subsection{Topological "order parameter" - Spin Chern number}

When $D\neq 0,$ one can not use the $Z_{2}$ topological number to classify
the topological insulators. However $S_{z}$ is still a good quantum number,
the Hamiltonian is decoupled for each spin component, then the spin Chern
number is suitable to characterize different quantum states. The Chern
number of each spin component is defined as $Q_{\uparrow ,\downarrow }.$

Now we write the Hamiltonian of up-spin and down-spin fermions into
\begin{equation}
H=\sum_{k}\Psi _{k\uparrow }^{\dag }\mathbf{N}_{\uparrow }(k)\Psi
_{k\uparrow }+\sum_{k}\Psi _{k\downarrow }^{\dag }\mathbf{N}_{\downarrow
}(k)\Psi _{k\downarrow ,}
\end{equation}%
where $\Psi _{k\uparrow }=(\Psi _{k,A\uparrow },\Psi _{k,B\uparrow })^{T}$
and $\Psi _{k\downarrow }=(\Psi _{k,A\downarrow },\Psi _{k,B\downarrow
})^{T} $. In above Hamiltonian, we have define
\begin{eqnarray}
\mathbf{N}_{\uparrow }(k) &=&N_{\uparrow ,1}^{1}\tau ^{x}+N_{\uparrow
,2}\tau ^{y}+N_{\uparrow ,3}\tau ^{z},\text{ } \\
\mathbf{N}_{\downarrow }(k) &=&N_{\downarrow ,1}^{1}\tau ^{x}+N_{\downarrow
,2}\tau ^{y}+N_{\downarrow ,3}\tau ^{z},  \notag
\end{eqnarray}%
where
\begin{eqnarray}
N_{\uparrow ,3} &=&-1-2t[\cos k_{x}+\cos k_{y}]-D, \\
N_{\uparrow ,2} &=&-\sin k_{x},\text{ }N_{\uparrow ,1}=\sin k_{y},  \notag
\end{eqnarray}%
and
\begin{eqnarray}
N_{\downarrow ,3} &=&-1-2t[\cos k_{x}+\cos k_{y}]+D, \\
N_{\downarrow ,2} &=&-\sin k_{x},\text{ }N_{\downarrow ,1}=-\sin k_{y}.
\notag
\end{eqnarray}%
$\tau ^{x,y,z}$ are Pauli matrices.

Now two Chern numbers $Q_{\uparrow }$ and $Q_{\downarrow }$ for up-spin
particle and down-spin particle are obtained as\cite{qi1},
\begin{equation}
Q_{\uparrow }=\frac{1}{8\pi ^{2}}\int_{\Omega }d^{2}k[\mathbf{n}_{\uparrow
}\cdot \partial _{x}\mathbf{n}_{\uparrow }\times \partial _{y}\mathbf{n}%
_{\uparrow }],
\end{equation}%
and
\begin{equation}
Q_{\downarrow }=\frac{1}{8\pi ^{2}}\int_{\Omega }d^{2}k[\mathbf{n}%
_{\downarrow }\cdot \partial _{x}\mathbf{n}_{\downarrow }\times \partial _{y}%
\mathbf{n}_{\downarrow }].
\end{equation}%
Here $\mathbf{n}_{\uparrow }$ and $\mathbf{n}_{\downarrow }$ are defined as $%
\mathbf{n}_{\uparrow }=\frac{\mathbf{N}_{\uparrow }}{|\mathbf{N}_{\uparrow }|%
}$ and $\mathbf{n}_{\downarrow }=\frac{\mathbf{N}_{\downarrow }}{|\mathbf{N}%
_{\downarrow }|}$, respectively. $\Omega $ is the area of Brillouin zone.

In the QSH state, for up-spin, the Chern number is $Q_{\uparrow }=1$; for
down-spin, it is\ $Q_{\downarrow }=-1$. So the total Chern number is zero $%
Q=Q_{\uparrow }+Q_{\downarrow }=0$. However there is nonzero spin Chern
number $Q_{s}=Q_{\uparrow }-Q_{\downarrow }=2$. In the QAH state, for
up-spin, the Chern number is $Q_{\uparrow }=1$; for down-spin, it is\ $%
Q_{\downarrow }=0$. So the total Chern number is obtained as $Q=Q_{\uparrow
}+Q_{\downarrow }=1$ and spin Chern number $Q_{s}=Q_{\uparrow
}-Q_{\downarrow }=1$. In NI state, the Chern number is zero for each spin
component, $Q=Q_{\uparrow }=Q_{\downarrow }=0$. From the results of the
Chern number of different quantum states, one may imagine that the QAH state
may be considered as \emph{half} of the QSH state. This statement is also
supported by the following calculations of other topological parameters.

In order to give a clear comparison of the total Chern number and the spin
Chern number in different quantum states, we list the results in Table.1.%
\begin{equation*}
\begin{tabular}[t]{|c|c|c|c|}
\hline
& QSH & QAH & \ NI \\ \hline
Total Chern number & $0$ & $1$ & $0$ \\ \hline
Spin Chern number & $2$ & $1$ & $0$ \\ \hline
\end{tabular}%
\end{equation*}%
From the Table. 1, one can see obviously that the spin Chern number can be
regarded as the topological invariant to characterize the topological
insulators and their quantum phase transitions for the Hamiltonian with $%
S^{z}$-conservation\cite{sheng,pro}.

\subsubsection{Edge states}

In this part, we calculate the edge states in different states.

From the relationship between the spin Chern number and the number of the
(chiral) edge states for each spin component $n_{\uparrow ,\downarrow },$ we
have%
\begin{eqnarray}
n_{\uparrow } &=&\frac{1}{8\pi ^{2}}\int_{\Omega }d^{2}k\text{ }\mathbf{n}%
_{\uparrow }\cdot \partial _{x}\mathbf{n}_{\uparrow }\times \partial _{y}%
\mathbf{n}_{\uparrow }, \\
n_{\downarrow } &=&\frac{1}{8\pi ^{2}}\int_{\Omega }d^{2}k\text{ }\mathbf{n}%
_{\downarrow }\cdot \partial _{x}\mathbf{n}_{\downarrow }\times \partial _{y}%
\mathbf{n}_{\downarrow }.  \notag
\end{eqnarray}%
Different Chern numbers also characterize different edge states. In the QSH
state, the total number of the edge states is obtained as \cite{qi1}
\begin{eqnarray}
n &=&n_{\uparrow }+n_{\downarrow }=0, \\
n_{s} &=&n_{\uparrow }-n_{\downarrow }=2.  \notag
\end{eqnarray}%
Thus there are two edge states with opposite chirality. In the QAH\ state,
there is only one edge state $n=n_{\uparrow }+n_{\downarrow }=1.$ The number
of edge states in different phases are plotted in Table.2.
\begin{equation*}
\begin{tabular}[t]{|c|c|c|c|}
\hline
& QSH & QAH & \ NI \\ \hline
Edge state's number & $2$ & $1$ & $0$ \\ \hline
\end{tabular}%
\end{equation*}

The numerical results of the edge states in different states are shown in
Fig.2.
\begin{figure}[tbph]
\includegraphics[width = 8.0cm]{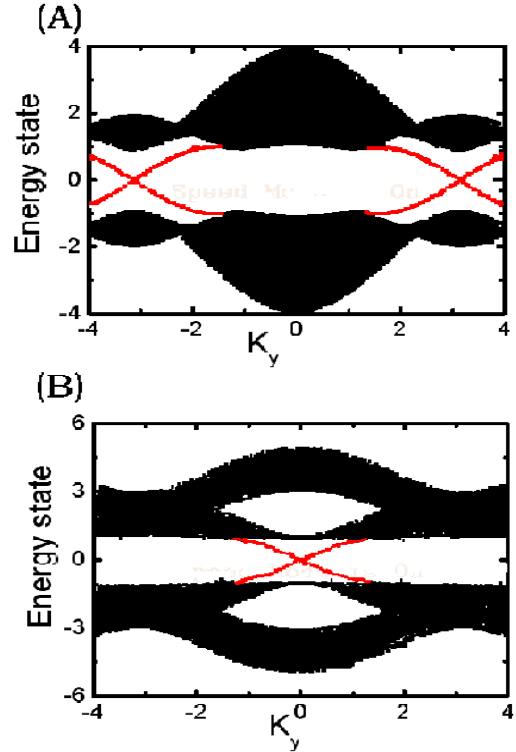}
\caption{(A): Energy spectrum of the QSH state when the open boundary
condition is imposed in x-direction. The parameters are $D=0,$ $t=0.75,$ $%
\protect\mu =2.$ (B): Energy spectrum of the QAH state when the open
boundary condition is imposed in x-direction. The parameters are $D=2,$ $%
t=0.5,$ $\protect\mu =2.$}
\end{figure}
From the Table.2, one can see that it is the spin Chern number that
determines the number of the edge states in different phases.

\subsubsection{Quantized Hall conductivity}

Next, we study the quantized Hall conductivity of different states.

We could calculate the charge Hall conductance $\sigma _{xy}$ and the
spin-Hall conductance $\sigma _{s,xy}$ for this model in different phases
using the standard representation\cite{qi1}
\begin{equation}
\sigma _{xy}=\sigma _{\uparrow ,xy}+\sigma _{\downarrow ,xy}.
\end{equation}%
and
\begin{equation}
\sigma _{s,xy}=\sigma _{\uparrow ,xy}-\sigma _{\downarrow ,xy.}
\end{equation}%
where%
\begin{eqnarray}
\sigma _{\uparrow ,xy} &=&\frac{e^{2}}{8\pi ^{2}h}\int_{\Omega }d^{2}k\text{
}\mathbf{n}_{\uparrow }\cdot \partial _{x}\mathbf{n}_{\uparrow }\times
\partial _{y}\mathbf{n}_{\uparrow }, \\
\sigma _{\downarrow ,xy} &=&\frac{e^{2}}{8\pi ^{2}h}\int_{\Omega }d^{2}k%
\text{ }\mathbf{n}_{\downarrow }\cdot \partial _{x}\mathbf{n}_{\downarrow
}\times \partial _{y}\mathbf{n}_{\downarrow }.  \notag
\end{eqnarray}

Here one can see that the physical consequence of the edge states is just
the quantum Hall conductance in our case. In the QSH state, the quantum Hall
conductance is obtained as
\begin{equation}
\sigma _{\uparrow ,xy}=\frac{e^{2}}{h},\text{ }\sigma _{\downarrow ,xy}=-%
\frac{e^{2}}{h}.
\end{equation}%
Thus there is no charge Hall conductance, $\sigma _{xy}=\sigma _{\uparrow
,xy}+\sigma _{\downarrow ,xy}=0$. Instead, the spin-Hall conductance is
non-zero, $\sigma _{s,xy}=\sigma _{\uparrow ,xy}-\sigma _{\downarrow ,xy}=%
\frac{2e^{2}}{h}.$ In the QAH\ state, the quantum Hall conductance is
obtained as
\begin{equation}
\sigma _{\uparrow ,xy}=\frac{e^{2}}{h},\text{ }\sigma _{\downarrow ,xy}=0.
\end{equation}%
Thus both charge Hall conductance and spin-Hall conductance in QAH state are
half of the spin-Hall conductance in QSH state
\begin{eqnarray}
\sigma _{xy} &=&\sigma _{\uparrow ,xy}+\sigma _{\downarrow ,xy}=\frac{e^{2}}{%
h},\text{ } \\
\sigma _{s,xy} &=&\sigma _{\uparrow ,xy}-\sigma _{\downarrow ,xy}=\frac{e^{2}%
}{h}.  \notag
\end{eqnarray}%
It is pointed out that the spin-Hall conductance is still well defined due
to $S_{z}$-conservation in the QAH state.

The spin Hall conductance in different states were shown in Fig.3. The
x-axis corresponded to the horizontal dash dot line in Fig.1. We can see
that the spin Hall conductance is $0$, $\frac{e^{2}}{h}$ and $\frac{2e^{2}}{h%
}$ in NI, QAH and QSH respectively. The spin Hall conductance changed from 0
to $\frac{e^{2}}{h}$ at the phase transition point $(t=0.125)$ and changed
from $\frac{e^{2}}{h}$ to $\frac{2e^{2}}{h}$ at the phase transition point $%
(t=0.375)$. In this sense, the TQPTs between QSH, QAH and NI are similar to
the quantum Hall plateau transition from Dirac fermion in Ref.\cite{lu}.

\begin{figure}[tbph]
\includegraphics[width = 8.0cm]{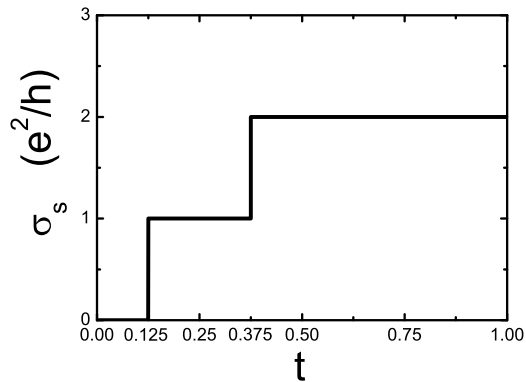}
\caption{The spin Hall conductance with respected to the three phases: NI,
QAH and QSH. The x-axis corresponded to the horizontal dash dot line in
Fig.1. }
\end{figure}

The clearly comparison of charge-Hall conductance and spin-Hall conductance
in different states are shown in Table.3
\begin{equation*}
\begin{tabular}[t]{|c|c|c|c|}
\hline
& QSH & QAH & \ NI \\ \hline
Charge Hall conductance ($\frac{e^{2}}{h}$) & $0$ & $1$ & $0$ \\ \hline
Spin Hall conductance ($\frac{e^{2}}{h}$) & $2$ & $1$ & $0$ \\ \hline
\end{tabular}%
\end{equation*}%
This table indicates that spin Chern number and total Chern number determine
charge Hall conductance and the spin-Hall conductance in different states,
respectively.

\subsubsection{Induced quantum numbers on a $\protect\pi $-flux}

In addition, we calculate the induced quantum number on a $\pi $-flux in
different quantum states. A $\pi $-flux denotes half a flux quantum $(\frac{1%
}{2}\Phi _{0},$ $\Phi _{0}=\frac{hc}{e})$ on one plaquette of the square
lattice which is shown in Fig.4.

\begin{figure}[tbph]
\includegraphics[width = 5.0cm]{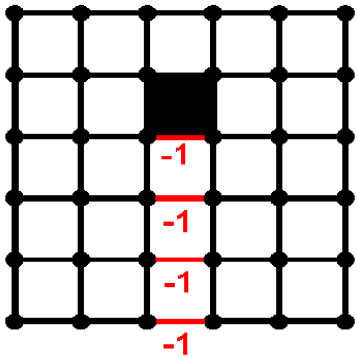}
\caption{Illustration of a $\protect\pi $-flux in a two-dimensional square
lattice ( the gray area ) at the end of a string of minus signs ($-1$) of
links.}
\end{figure}

We calculate the induced quantum number on a $\pi $-flux in the QSH state
firstly. In the QSH state, the $\pi $-flux defects defined above may be
considered as the Jackiw-Rebbi soliton\cite{jackiw}. From the numerical
results, we find that there are two zero modes of a single $\pi $-flux
defect. So such a two dimensional point defect possesses four soliton
states, two for each spin (occupied and non-occupied) which are denoted by
\begin{equation}
\mid \uparrow _{+}\rangle \otimes \mid \downarrow _{+}\rangle ,\text{ }\mid
\uparrow _{-}\rangle \otimes \mid \downarrow _{-}\rangle ,
\end{equation}%
and%
\begin{equation}
\mid \uparrow _{-}\rangle \otimes \mid \downarrow _{+}\rangle \text{ }\mid
\uparrow _{+}\rangle \otimes \mid \downarrow _{-}\rangle .
\end{equation}%
Here $\mid \uparrow _{-}\rangle $ and $\mid \downarrow _{-}\rangle $ are the
empty states of the zero modes $\Psi _{\uparrow }^{0}(r)$ and $\Psi
_{\downarrow }^{0}(r).$ Around a $\pi $ flux, the fermionic operators are
expanded as
\begin{eqnarray}
\hat{\Psi}_{\alpha }(r,t) &=&\sum_{k\neq 0}\hat{b}_{\alpha
k}e^{-iE_{k}t}\Psi _{\alpha k}(r) \\
&&+\sum_{k\neq 0}\hat{d}_{\alpha k}^{\dagger }e^{iE_{k}t}\Psi _{\alpha
k}^{\dagger }(r)+\hat{a}_{\alpha }^{0}\Psi _{\alpha }^{0}(r),\text{ }  \notag
\end{eqnarray}%
where $\hat{b}_{\alpha k}$ and $\hat{d}_{\alpha k}^{\dagger }$ are operators
of $k\neq 0$ modes that are irrelevant to the soliton states discussed
below. $\hat{a}_{\alpha }^{0}$ are annihilation operators of zero modes and $%
\alpha $ is the spin index. Thus we have the relationships as
\begin{eqnarray}
\hat{a}_{\uparrow }^{0} &\mid &\uparrow _{+}\rangle =\mid \uparrow
_{-}\rangle ,\text{ }\hat{a}_{\uparrow }^{0}\mid \uparrow _{-}\rangle =0,%
\text{ }  \label{a} \\
\hat{a}_{\downarrow }^{0} &\mid &\downarrow _{+}\rangle =\mid \downarrow
_{-}\rangle ,\text{ }\hat{a}_{\downarrow }^{0}\mid \downarrow _{-}\rangle =0.%
\text{ }  \notag
\end{eqnarray}

We define the induced fermion number operators of the soliton states, $\hat{N%
}_{\alpha ,F}$ with
\begin{equation}
\hat{N}_{\alpha ,F}\equiv (\hat{a}_{\alpha }^{0})^{\dagger }\hat{a}_{\alpha
}^{0}+\sum \limits_{k\neq 0}(\hat{b}_{\alpha k}^{\dagger }\hat{b}_{\alpha k}-%
\hat{d}_{\alpha k}^{\dagger }\hat{d}_{\alpha k})-\frac{1}{2},  \notag
\end{equation}%
From the relation in Eq.(\ref{a}), we find that $\mid \uparrow _{\pm
}\rangle $ or $\mid \downarrow _{\pm }\rangle $ have eigenvalues of $\pm
\frac{1}{2}$ of the induced fermion number operators,
\begin{eqnarray}
\hat{N}_{\uparrow ,F}| &\uparrow &_{\pm }\rangle =\pm {\frac{1}{2}}|\uparrow
_{\pm }\rangle ,\text{ }\hat{N}_{\uparrow ,F}|\downarrow _{\pm }\rangle =0,%
\text{ }  \label{half} \\
\hat{N}_{\downarrow ,F}| &\downarrow &_{\pm }\rangle =\pm {\frac{1}{2}}%
|\downarrow _{\pm }\rangle ,\text{ }\hat{N}_{\downarrow ,F}|\uparrow _{\pm
}\rangle =0.\text{ }  \notag
\end{eqnarray}%
From $\hat{N}_{\alpha ,F},$ we can define two induced quantum number
operators, the total induced fermion number operator $\hat{N}_{F}=\hat{N}%
_{\uparrow ,F}+\hat{N}_{\downarrow ,F}$ and the induced quantum spin
operator, $\hat{S}^{z}=\frac{1}{2}(\hat{N}_{\uparrow ,F}-\hat{N}_{\downarrow
,F}).$

Then we calculate two induced quantum numbers defined above. For an
insulator state, the ground states are the two degenerate soliton states
denoted by $\mid \uparrow _{-}\rangle \otimes \mid \downarrow _{+}\rangle $
and $\mid \uparrow _{+}\rangle \otimes \mid \downarrow _{-}\rangle $. One
can easily check that the total induced fermion number on the solitons is
zero from the cancelation effect between different spin components
\begin{equation}
\hat{N}_{F}\mid \uparrow _{-}\rangle \otimes \mid \downarrow _{+}\rangle =%
\hat{N}_{F}\mid \uparrow _{+}\rangle \otimes \mid \downarrow _{-}\rangle =0.
\end{equation}%
On the other hand, there exists \emph{a\ spin-}$\frac{1}{2}$\emph{\ moment}
on the soliton states $\mid \uparrow _{-}\rangle \otimes \mid \downarrow
_{+}\rangle $ and $\mid \uparrow _{+}\rangle \otimes \mid \downarrow
_{-}\rangle ,$
\begin{eqnarray}
\hat{S}^{z} &\mid &\uparrow _{-}\rangle \otimes \mid \downarrow _{+}\rangle =%
\frac{1}{2}\mid \uparrow _{-}\rangle \otimes \mid \downarrow _{+}\rangle , \\
\hat{S}^{z} &\mid &\uparrow _{+}\rangle \otimes \mid \downarrow _{-}\rangle
=-\frac{1}{2}\mid \uparrow _{+}\rangle \otimes \mid \downarrow _{-}\rangle .
\notag
\end{eqnarray}%
The induced spin moment may be straightforwardly obtained by combining the
definition of $\hat{S}^{z}$ and Eq. (\ref{half}) together.

The four different ways of occupation these two zero modes give rise to four
different type of fluxons with the following quantum numbers: ($charge=0,$ $%
spin=\pm \frac{1}{2}$) and ($charge=\pm 1,$ $spin=0$). The results is
consistent to that there exist spin-charge separated solitons in the
presence of $\pi $-flux with induced quantum numbers\cite{qi,ran,kou1}.

The charge density of a fluxon in the QSH state on a $24\times 24$ lattice $%
\Psi _{\uparrow }^{0}(r)\Psi _{\uparrow }^{0}(r)+\Psi _{\downarrow
}^{0}(r)\Psi _{\downarrow }^{0}(r)$ is shown in Fig.5. The zero mode is
localized around the defect center within a length-scale $\sim m^{-1}.$ Here
$m$ is the mass gap of the fermions.

\begin{figure}[tbph]
\includegraphics[width = 10.0cm]{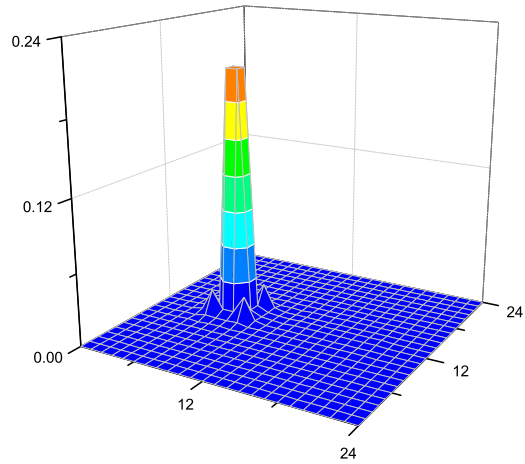}
\caption{The charge density of a fluxon in the QSH state on a $24\times 24$
lattice with periodic boundary condition. The charge bound to the defect is $%
\pm e.$}
\end{figure}

Secondly, we study the zero modes and the induced quantum numbers on a $\pi $%
-flux in the QAH state. In the QAH state, there is only one zero mode of a
single $\pi $-flux defect. The charge densities $\Psi _{\uparrow
}^{0}(r)\Psi _{\uparrow }^{0}(r)+\Psi _{\downarrow }^{0}(r)\Psi _{\downarrow
}^{0}(r)$ of a pair of defects in the QAH state on a $24\times 24$ lattice
with periodic boundary condition was shown in Fig.6. For the two defects,
the two zero modes slightly split due to tunneling effect between them.

\begin{figure}[tbph]
\includegraphics[width = 10.0cm]{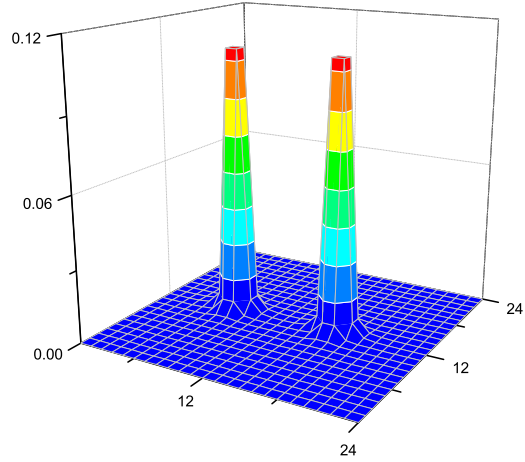}
\caption{The charge densities of a pair of defects in the QAH state on a $%
24\times 24$ lattice with periodic boundary condition. The charge bound to
the defect is $\pm \frac{e}{2}.$}
\end{figure}

Thus there are two soliton states, two for up spin particles (occupied and
non-occupied) which are denoted by $\mid \uparrow _{+}\rangle $ and $\mid
\uparrow _{-}\rangle .$ In contrast, there is no zero mode and soliton
states of the $\pi $-flux defect for down spin particles. Similarly we can
obtain the eigenvalues of the total induced fermion number operator $\hat{N}%
_{F}$,
\begin{equation}
\hat{N}_{F}|\uparrow _{\pm }\rangle =\hat{N}_{\uparrow ,F}|\uparrow _{\pm
}\rangle =\pm {\frac{1}{2}}|\uparrow _{\pm }\rangle .
\end{equation}%
And the induced quantum spin number on the soliton states $\mid \uparrow
_{\pm }\rangle $ is obtained as
\begin{eqnarray}
\hat{S}^{z} &\mid &\uparrow _{\pm }\rangle =\frac{1}{2}(\hat{N}_{\uparrow
,F}-\hat{N}_{\downarrow ,F})|\uparrow _{\pm }\rangle  \notag \\
&=&\frac{1}{2}\hat{N}_{\uparrow ,F}|\uparrow _{\pm }\rangle =\pm \frac{1}{4}%
\mid \uparrow _{\pm }\rangle .
\end{eqnarray}%
The occupation (or unoccupation) of this zero mode lead to $N_{F}=\pm \frac{e%
}{2}$ charge and $S^{z}=\frac{1}{4}$ bound to the defect\cite{franz}. Due to
$S^{z}$-conservation for the QAH state, the induced quantum spin number $%
S^{z}=\frac{1}{4}$ is also well defined.

Thirdly, in the NI state, there is no zero mode on a single $\pi $-flux
defect. As a result, all induced quantum numbers are zero.\

Therefore, in this part we find that the zero mode's on a $\pi $-flux may
also be used to distinguish different quantum phases. These induced quantum
numbers are shown in Table.4.
\begin{equation*}
\begin{tabular}[t]{|c|c|c|c|}
\hline
& QSH & QAH & \ NI\  \\ \hline
Zero mode's number & $2$ & $1$ & $0$ \\ \hline
$N_{F}$ & $0$ & $\frac{1}{2}$ & $0$ \\ \hline
$S^{z}$ & $\frac{1}{2}$ & $\frac{1}{4}$ & $0$ \\ \hline
\end{tabular}%
\end{equation*}

Finally we give a summary. When $D\neq 0,$ because $S_{z}$ is a good quantum
number, we can use the spin Chern number $Q_{s}$ to characterize different
quantum states as shown in the Table.5.

\begin{equation*}
\begin{tabular}[t]{|c|c|c|c|}
\hline
& QSH & QAH & \ NI\  \\ \hline
Spin Chern number $Q_{s}$ & $2$ & $1$ & $0$ \\ \hline
Edge state's number $n_{s}$ & $2$ & $1$ & $0$ \\ \hline
Zero mode's number & $2$ & $1$ & $0$ \\ \hline
\end{tabular}%
\end{equation*}%
For the TQPTs from one quantum state to another, the spin Chern number $%
Q_{s} $ will jump.

\subsection{Universal critical behavior of TQPTs}

\begin{figure}[tbph]
\includegraphics[width = 8.0cm]{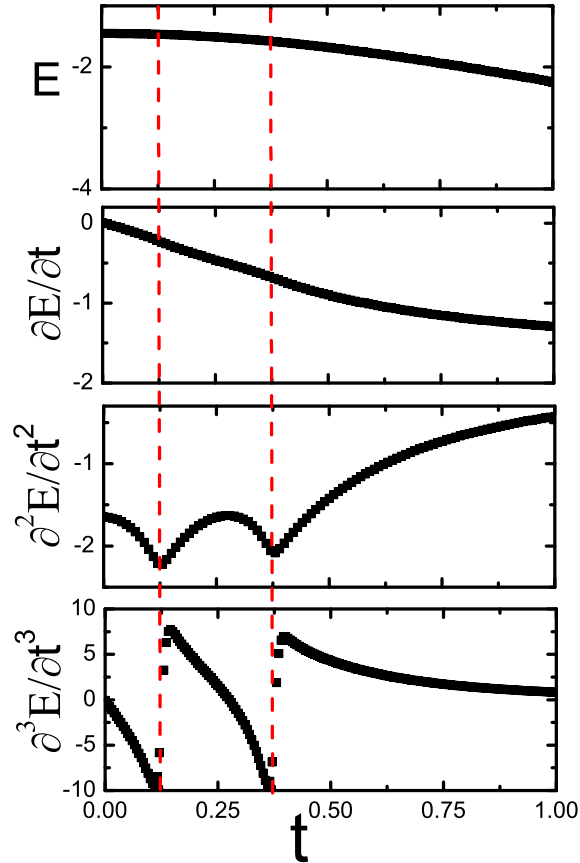}
\caption{The ground-state energy and its first, second and third derivatives
with respect to $t$, where $D=0.5$ (which corresponds to the horizontal dash
dot line in Fig.1). It is clear that $E,\frac{\partial E}{\partial t}$ and $%
\frac{\partial ^{2}E}{\partial t^{2}}$ are continuous functions, but $\frac{%
\partial ^{3}E}{\partial t3}$ is discontinuous at the transition points $%
t=0.125$ and $0.375.$}
\end{figure}

Another feature of the TQPTs is the non-analyticity of the ground-state
energy. The ground-state energy is defined as

\begin{equation}
E=-\sum_{k}(\varepsilon _{+(k)}+\varepsilon _{-(k)})=-\frac{S}{4\pi ^{2}}%
\underset{BZ}{\int }(\varepsilon _{+(k)}+\varepsilon _{-(k)})d^{2}k,
\end{equation}%
where, $BZ$ denotes the first Brillouin zone and

\begin{equation}
\varepsilon _{\pm (k)}=\sqrt{\sin ^{2}k_{y}+\sin ^{2}k_{x}+(\frac{\mu }{2}%
+2t(\cos k_{x}+\cos k_{y})\pm D)^{2}}.
\end{equation}%
Here $S$ is the area of the system. As an example, we show the TQPT of the
model along the line with fixing $D=0.5$ in Fig.1. As illustrated in Fig.7,
the ground state energy and its first and second derivatives are continuous
for arbitrary $t$, while its third derivative is non-analytic at points $%
t=0.125$ and $t=0.375$, corresponding to phase transitions NI -- QAH and QAH
-- QSH. It means that the TQPTs are third order. Similar third order TQPTs
have been pointed out in other systems where the nodal fermion appears\cite%
{cai}.

Let us explain why TQPTs are always third order. The energy dispersions near
the quantum phase transitions are shown in Fig.8 and Fig.9. On the line with
fixing $D=0.5,$ one can see that there is one nodal point with zero energy
in the Brillouin zone, $(\pi ,$ $\pi ),$ at the quantum phase transition
between QSH state and QAH state. Thus the TQPT is dominated by nodal Dirac
fermionic excitations at $(\pi ,$ $\pi )$. For the TQPT between QAH state
and NI state on the line with fixing $D=0.5$, the nodal fermion is also
pinned at the point $(\pi ,$ $\pi )$.
\begin{figure}[tbph]
\includegraphics[width = 8.0cm]{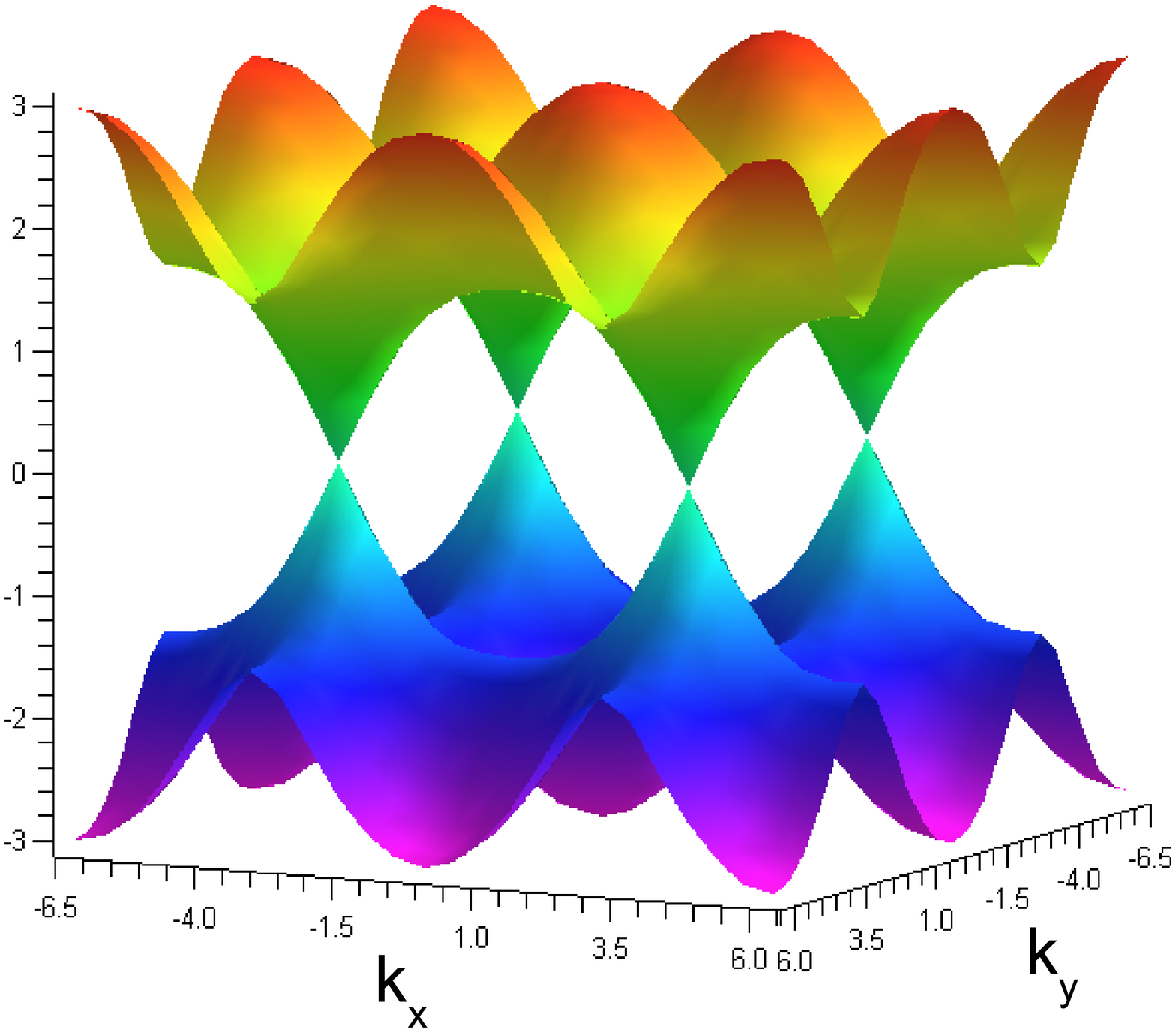}
\caption{The energy dispersion near the point of phase transition between
QSH and QAH. ($k_{x},$ $k_{y}:(-2\protect\pi ,$ $2\protect\pi )$)}
\end{figure}

\begin{figure}[tbph]
\includegraphics[width = 8.0cm]{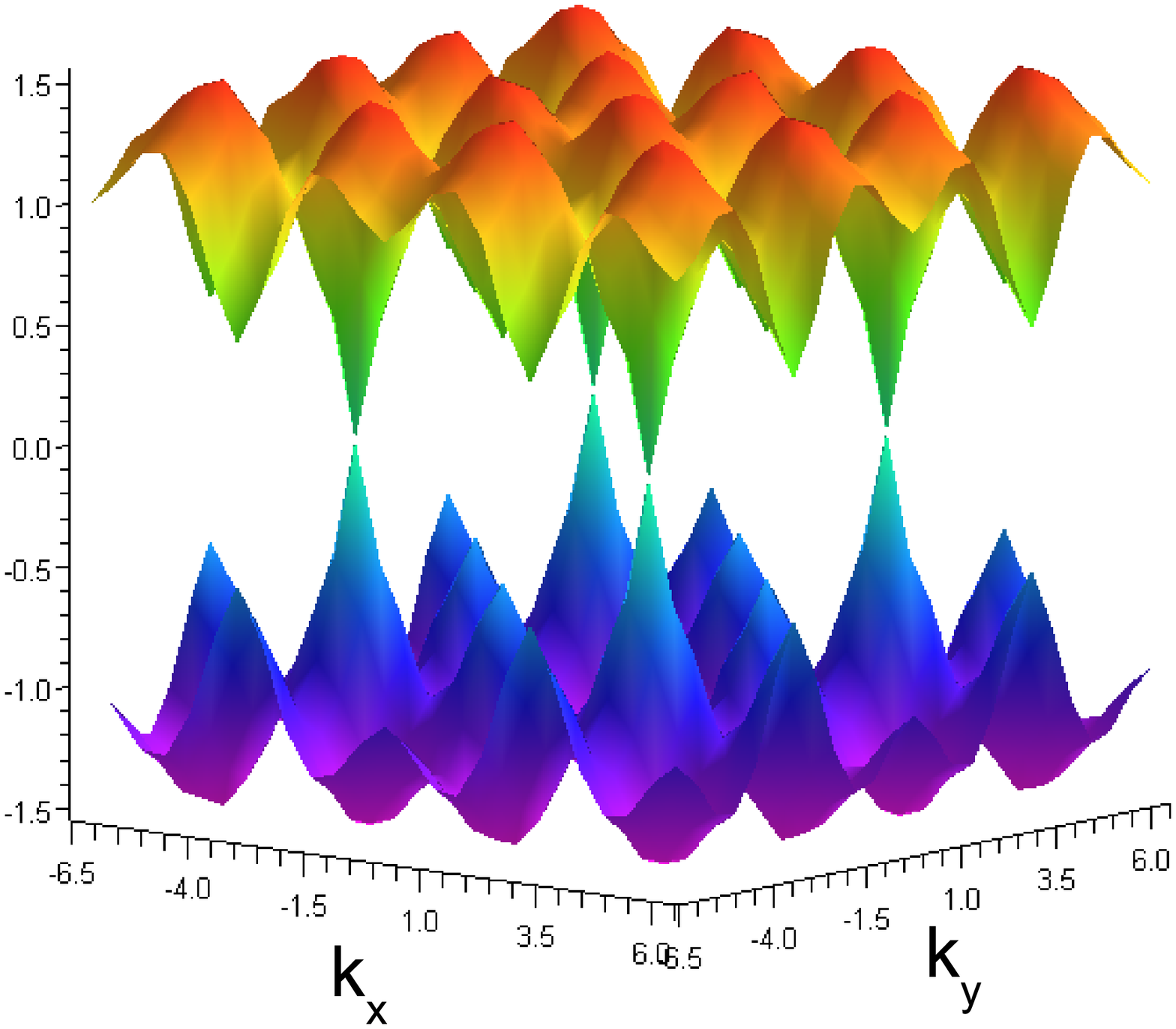}
\caption{The energy dispersion near the point of phase transition between NI
and QAH. ($k_{x},$ $k_{y}:(-2\protect\pi ,$ $2\protect\pi )$)}
\end{figure}

Furthermore, near the critical points of the TQPTs, we get the low energy
effective Hamiltonian near the point $(\pi ,$ $\pi ),$
\begin{equation}
H_{1,\uparrow }\rightarrow m_{\uparrow }\tau ^{z}+k_{y}\tau ^{x}+k_{x}\tau
^{y}
\end{equation}%
and%
\begin{equation}
H_{1,\downarrow }\rightarrow m_{\downarrow }\tau ^{z}+k_{y}\tau
^{x}-k_{x}\tau ^{y}
\end{equation}%
where $m_{\uparrow }=-1+4t-D$ and $m_{\downarrow }=-1+4t+D$ are the masses
of the electrons with up-spin and down-spin, respectively. One can see that
crossing the critical point $D=0.5$, $t=0.125$, the mass of the electrons
with down-spin changes sign, $\frac{m_{\downarrow }}{\left\vert
m_{\downarrow }\right\vert }\rightarrow -\frac{m_{\downarrow }}{\left\vert
m_{\downarrow }\right\vert }.$ Consequently, the Chern number of the
electrons with down-spin $Q_{\downarrow }$ changes from $0$ to $1$ that
corresponds to the TQPTs from NI state ($Q_{\uparrow }=Q_{\downarrow }=0$)
to QAH state ($Q_{\uparrow }=0,$ $Q_{\downarrow }=1$). Similarly, crossing
the critical point $D=0.5$, $t=0.375$, the mass of the electrons with
up-spin changes sign, $\frac{m_{\uparrow }}{\left\vert m_{\uparrow
}\right\vert }\rightarrow -\frac{m_{\uparrow }}{\left\vert m_{\uparrow
}\right\vert }$, that leads to the Chern number of the electrons with
up-spin $Q_{\uparrow }$ changes from $0$ to $1$, corresponding to the TQPTs
from QAH state ($Q_{\uparrow }=0,$ $Q_{\downarrow }=1$) to QSH state ($%
Q_{\uparrow }=1,$ $Q_{\downarrow }=1$). For a special TQPT at ($D=0$), the
mass of the electrons with up-spin $m_{\uparrow }=-1+4t$ is equal to the
mass of electrons with down-spin $m_{\downarrow }=-1+4t.$ The changes of
mass sign will lead to the jumps of spin Chern number, $\Delta Q_{\uparrow
}=\Delta Q_{\downarrow }=1,$ that corresponds to a direct topological
quantum phase transition from NI state ($Q_{\uparrow }=Q_{\downarrow }=0$)
to QSH state ($Q_{\uparrow }=1,$ $Q_{\downarrow }=1$)\cite{shu}. Fig.10 is a
scheme to illustrate the relationship between the jump of the spin Chern
number and the change of mass sign.

\begin{figure}[tbph]
\includegraphics[width = 12.0cm]{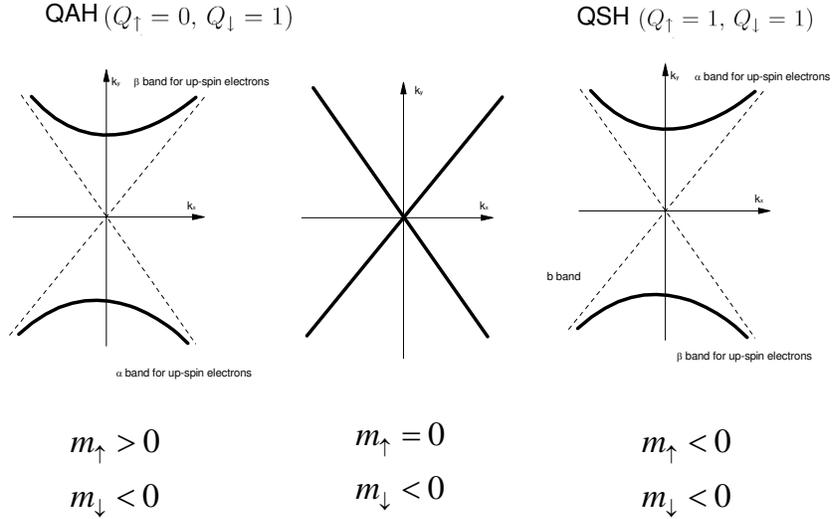}
\caption{Illustration of the relationship between the jump of the spin Chern
number and the change of mass sign.}
\end{figure}

On the other hand, due to the existence of the nodal fermion, the TQPTs with
mass sign changes ($m_{\uparrow ,\downarrow }\rightarrow -m_{\uparrow
,\downarrow }$) are always third order. One may use the low energy
approximation, $E(k)=\sqrt{k^{2}+m_{\uparrow ,\downarrow }^{2}},$ and then
get the ground state energy as
\begin{equation}
E(m_{\uparrow ,\downarrow })\sim -|m_{\uparrow ,\downarrow }|^{3}+\mathrm{%
const}_{\uparrow ,\downarrow }
\end{equation}%
where $\mathrm{const}_{\uparrow ,\downarrow }$ is a constant\cite{cai}. It
is obvious that the third order derivative of $E(m_{\uparrow ,\downarrow })$
to $m_{\uparrow ,\downarrow }$ is discontinuous at the point $m_{\uparrow
,\downarrow }=0$.

In brief, for the TQPTs, NI -- QAH and QAH -- QSH, the changes of the mass
sign $m_{\uparrow ,\downarrow }\rightarrow -m_{\uparrow ,\downarrow }$
reflect the jumps of the (spin) Chern numbers ($\Delta Q_{s}\neq 0$). As a
result, third order phase transitions is\ a universal feature of the TQPTs.

\subsection{Finite temperature properties}

\begin{figure}[tbph]
\includegraphics[width = 10.0cm]{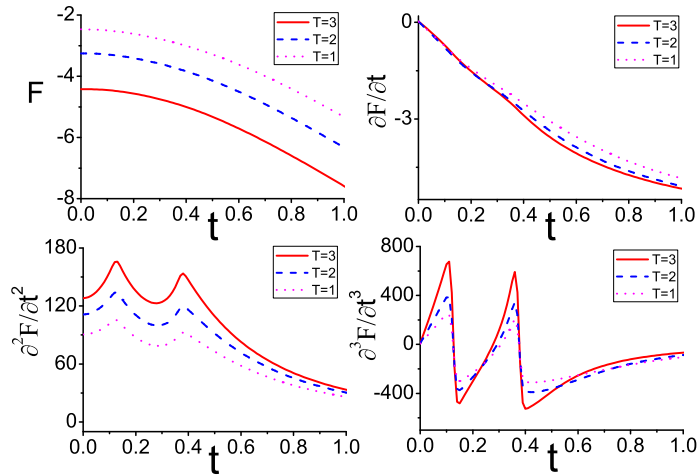}
\caption{The free energy and its first, second and third derivatives with
respect to $t$, where $D=0.5$ (which corresponds to the horizontal dash dot
line in Fig.1). It is clear that $F,$ $\frac{\partial F}{\partial t}$, $%
\frac{\partial ^{2}F}{\partial t^{2}}$ and $\frac{\partial ^{3}F}{\partial t3%
}$ are all analytic.}
\end{figure}

At finite temperature ($T\neq 0$), the free energy $F$ is defined by
\begin{equation}
F=-T\int \frac{d^{2}k}{(2\pi )^{2}}2\ln (\cosh (\varepsilon
_{+(k)}+\varepsilon _{-(k)})).
\end{equation}%
Here we set $k_{B}=1$. We calculate the derivatives of the free energy which
are shown in Fig.11. One can see that the free energy and its first, second,
third derivatives are all analytic without singularity at $T\neq 0$. The
results means that at finite temperature, there is no real phase transition,
instead one gets crossovers. This result is consistent with that of the TQPT
between QAH state and NI state in Ref.\cite{cai}.

\subsection{Stability of the TQPTs}

Finally we discuss the stability of the TQPTs. At the critical points, there
always exist nodal fermion with zero energy at $(\pi ,$ $\pi )$. Thus at the
critical points, we linearize the dispersion at $(\pi ,$ $\pi )$ and get a
low energy effective Dirac Hamiltonian of nodal fermions,%
\begin{eqnarray}
H_{1} &=&\sum_{k}\Psi _{k,\uparrow }^{\dagger }(\tau ^{x}k_{y}+\tau
^{y}k_{x})\Psi _{k,\uparrow }+\sum_{k}\Psi _{k,\downarrow }^{\dagger }(\tau
^{y}k_{x}-\tau ^{x}k_{y})\Psi _{k,\downarrow } \\
&=&\sum_{k}\Psi _{k}^{\dagger }\sigma _{z}\gamma _{i}k_{i}\Psi _{k},  \notag
\end{eqnarray}%
where the $\gamma $ matrices are defined by $\gamma _{x}=\tau ^{2}$ and $%
\gamma _{y}=\tau ^{1}$.

Then we consider effect of short range interactions. For example, one may
add an on-site four-fermi interaction $H_{int}=U(\Psi ^{\dagger }\Psi )^{2}$
to $H_{1}.$ Now we get an effective three dimensional Gross-Neveu model with
the Lagrangian\cite{gross}
\begin{equation}
L_{GN}=i\Psi ^{\dagger }\gamma _{\mu }\partial _{\mu }\Psi +U(\Psi ^{\dagger
}\Psi )^{2}.
\end{equation}%
It is known in large-N limit, the Callan-Symanzik function of $\beta (U)$ is
$\beta (U)=-\frac{\lambda U}{2\pi }+U$. For a small interaction, $%
U\rightarrow 0,$ the four-fermi interaction is irrelevant. That means the
TQPTs is stable against the small on-site four-fermi interaction. Similarly,
considering other types of short range interaction with $S_{z}$%
-conservation, one may get the same results: they are irrelevant and the
nodal fermions are stable.

\section{TQPTs with T-symmetry : $D=0,$ $R\neq 0$}

\subsection{Global phase diagram}

In this section, to learn the TQPT of topological insulators with
T-symmetry, $D=0,$ $R\neq 0,\ $we focus on $H_{2}=H_{0}+H_{R}$ and show the
effect of the Rashba term.

\begin{figure}[tbph]
\includegraphics[width = 10.0cm]{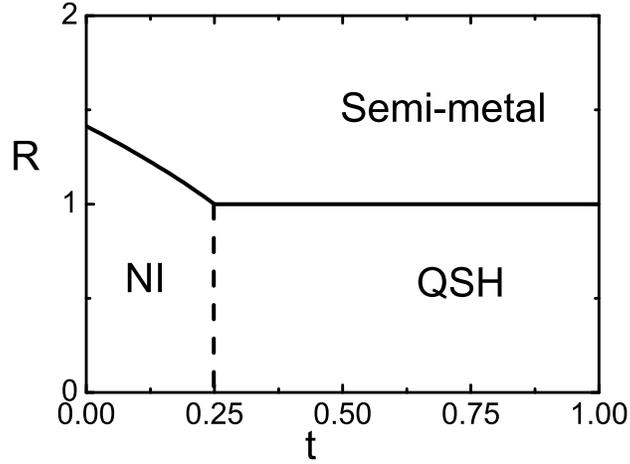}
\caption{The global phase diagram at $D=0$.}
\end{figure}

From the global phase diagram in Fig.12, one can see that there exist three
quantum phases : QSH state, NI state and gapless semi-metal. The solid line
divides the gapped phases and the gapless semi-metal. As shown in Fig.13,
when one fixes $t$ at $0.375$, the energy gap will decrease with the
increase of $R$ and close at $R=1.$ However, the phase boundary (the dashed
line in Fig.12) between the QSH state and NI state doesn't change by the
Rashba term.

\begin{figure}[tbph]
\includegraphics[width = 10.0cm]{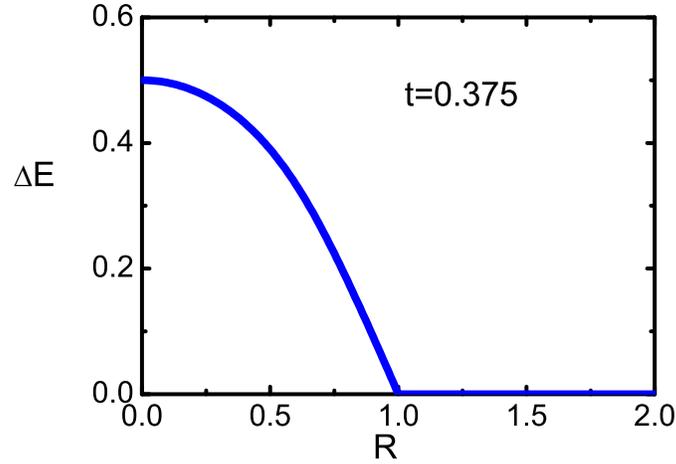}
\caption{The energy gap with respect to the Rashba term $R$ at $t=0.375$.}
\end{figure}

\subsection{Topological "order parameter" - $Z_{2}$ topological invariant}

To see the topological properties of the Hamiltonian with T-symmetry but
without $S_{z}$-conservation, we calculate the edge states and the zero
mode's number on a $\pi $-flux in different quantum states. In the QSH
state, from the numerical results, we find that there also exist two zero
modes of a single $\pi $-flux defect. The charge density of a fluxon in the
QSH state on a $24\times 24$ lattice is shown in Fig.14. In the NI state,
there is no such zero mode on a single $\pi $-flux defect. On the other
hand, the results of the edge states are shown in Fig.15, from which we find
that there are two edge states in QSH state, while in the NI state, there is
no edge state.\
\begin{figure}[tbph]
\includegraphics[width = 10.0cm]{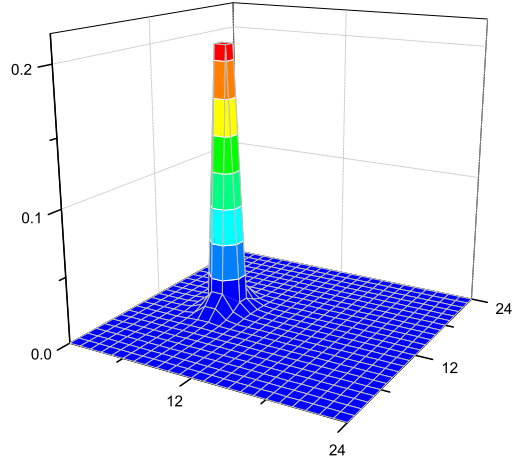}
\caption{The charge density of a fluxon in the QSH phase on a $24\times 24$
lattice with periodic boundary condition. The charge bound to the defect is $%
\pm e.$}
\end{figure}
\begin{figure}[tbph]
\includegraphics[width = 8.0cm]{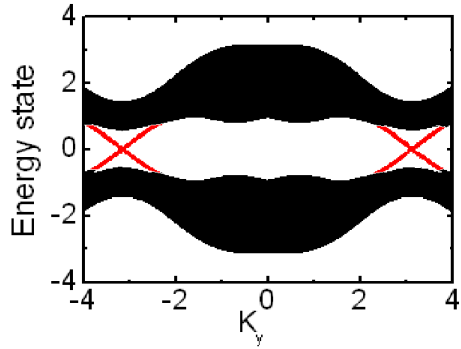}
\caption{Energy spectrum of the QSH state when the open boundary condition
is imposed in x-direction. The parameters are $R=0.4,$ $t=0.5,$ $\protect\mu %
=2.$ }
\end{figure}

Above results indicate that the T-symmetry indeed protects the topological
properties of the QSH state. Thus we calculate the $Z_{2}$ topological
invariant $(-1)^{\Delta }=\prod_{i=1}^{4}\prod_{m=1}^{N}\xi _{2m}(\vec{k}%
_{i})$ in different gapped quantum state. For the four high-symmetry points,
$(0,0)$, $(0,\pi )$, $(\pi ,0)$, $(\pi ,\pi )$, the eigenvalues below Fermi
surface $E_{F}$ of this system are
\begin{equation}
1+4t,\text{ }1,\text{ }1,\text{ }1-4t,  \label{z2}
\end{equation}%
respectively. TQPT occurs at $t=1/4$. For $t<1/4,$ the four eigenvalues are
all positive. Using the formula mentioned above, one can see that the $Z_{2}$
topological invariant $(-1)^{\Delta }$ is $1$ which means the system is in
the NI phase. For $t>1/4$, one eigenvalue is negative while the others are
all positive. It is obvious that the $Z_{2}$ topological invariant $%
(-1)^{\Delta }$ is $-1$ which means the system is in the QSH phase.

Table.6 shows the relationship between the $Z_{2}$ topological invariant $%
(-1)^{\Delta }$ and the topological properties of the quantum states,

\begin{equation*}
\begin{tabular}[t]{|c|c|c|}
\hline
$D=0,$ $R\neq 0$ & QSH & \ NI\  \\ \hline
$Z_{2}$ topological invariant $(-1)^{\Delta }$ & $-1$ & $1$ \\ \hline
Edge state's number & $2$ & $0$ \\ \hline
Zero mode's number on a $\pi $-flux & $2$ & $0$ \\ \hline
\end{tabular}%
\end{equation*}

\subsection{Universal critical behavior of TQPT}

To study the universal critical behavior of the TQPT\ between QSH state and
NI state without $S_{z}$-conservation, we calculate the derivatives of the
ground-state energy $E$. Here the ground-state energy becomes $%
E=-\sum_{k}(\varepsilon _{+(k)}+\varepsilon _{-(k)})$ where
\begin{equation}
\varepsilon _{\pm (k)}=\sqrt{\sin ^{2}k_{y}+(R\sqrt{(\sin ^{2}k_{x}+\sin
^{2}k_{y})}\pm \sqrt{\sin ^{2}k_{y}+(\frac{\mu }{2}+2t(\cos k_{x}+\cos
k_{y}))^{2}})^{2}}.
\end{equation}%
We show the TQPT along the line with fixing $R=0.2$. As illustrated in
Fig.16, the third derivative is non-analytic at points $t=0.25$,
corresponding to quantum phase transitions NI -- QSH. So the TQPT is also
third order.
\begin{figure}[tbph]
\includegraphics[width = 8.0cm]{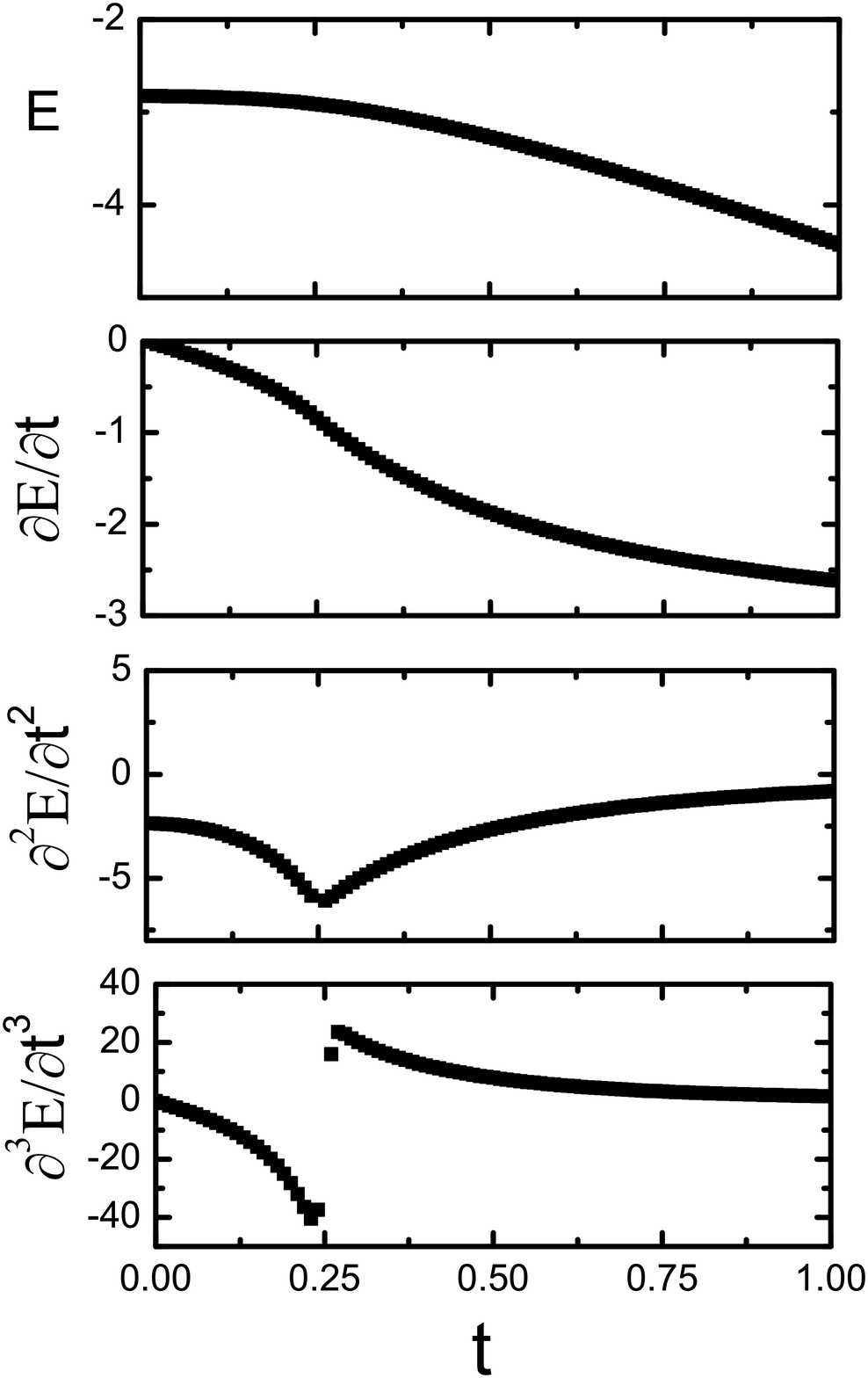}
\caption{The ground-state energy and its first, second and third derivatives
with respect to $t$, It is clear that $E,\frac{\partial E}{\partial t}$ and $%
\frac{\partial ^{2}E}{\partial t^{2}}$ are continuous functions, but $\frac{%
\partial ^{3}E}{\partial t3}$ is discontinuous at the transition point $%
t=0.25$.}
\end{figure}

The energy dispersion near the quantum phase transition is shown in Fig.17.
From the dispersion of the electrons near the critical points, we find that
the TQPT is also dominated by the nodal fermion near the point $(\pi ,$ $\pi
).$

\begin{figure}[tbph]
\includegraphics[width = 8.0cm]{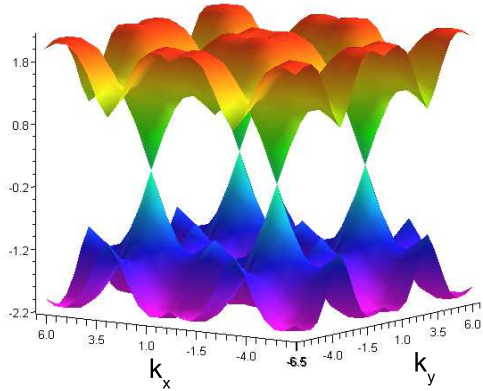}
\caption{The energy dispersion near the quantum phase transition. The
parameters are $t=0.25$ and $R=0.2.$}
\end{figure}

Let us explain why the TQPT between QSH state and NI\ state are also third
order. When $R\neq 0,$ there is no $S_{z}$-conservation and one cannot
define the spin Chern number. So we cannot use the jump of the Chern number
to understand the TQPT. However, from the definition of the $Z_{2}$
topological invariant, we find that near the TQPT, the mass signs of the two
kinds of low energy fermions ($m=-1+4t$) are determined by the $Z_{2}$
topological invariant $(-1)^{\Delta }$ (See Eq.\ref{z2}). Consequently, for
the TQPT between NI state ($(-1)^{\Delta }=1$) to QSH state ($(-1)^{\Delta
}=-1$), the sudden change of the $Z_{2}$ topological invariant will lead to
the change of mass sign of low energy fermions.

In brief, for the TQPT, NI -- QSH, the change of mass sign come from the
sudden change of the $Z_{2}$ topological invariant. As a result, the TQPT
between NI\ state and QSH state is always third order.

\section{TQPTs without T-symmetry and $S_{z}$-conservation : $D\neq 0,$ $%
R\neq 0$}

In this section, we study the TQPTs of topological insulators without
T-symmetry and $S_{z}$-conservation, $D\neq 0,$ $R\neq 0\ $based on $%
H_{3}=H_{0}+H_{D}+H_{R}.$

\begin{figure}[tbph]
\includegraphics[width = 10.0cm]{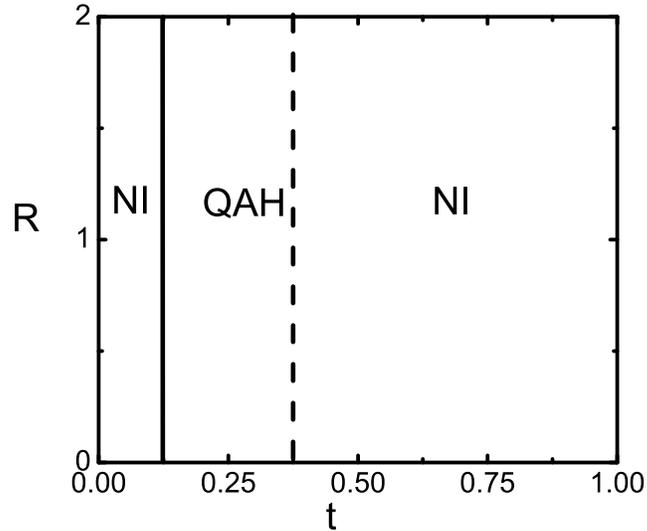}
\caption{The global phase diagram at $D=0.5.$}
\end{figure}

We take $D=0.5$ as an example (which corresponds to the horizontal dash dot
line in Fig.1). The results are illustrated in Fig.18. One can see that in
the phase diagram there exist three phases : two NI states and one QAH
state. The QAH state is robust while arbitrary small $R$ will destroy the
QSH state. To understand the disappearance of the QSH state, on the one hand
the evolution of the energy levels of zero modes on a $\pi $-flux with $R$
at the point $D=0.5,$ $t=0.75$ is plotted in Fig.19. The black bars denote
the continuum spectrum while the black triangles denote the energy levels of
zero modes on a $\pi $-flux. One can see that the degeneracy of zero modes
is lifted by the Rashba term. With increasing $R,$ the energy gap of the
continuum spectrum decreases, while\ the energy splitting of the zero modes $%
\Delta E$ increases. On the other hand, the evolution of edge states in QSH
state with the Rashba term is shown in Fig.20. When $R\neq 0$ and $D\neq 0,$
due to the hybridization of two edge states, the edge states in the QSH
state open an energy gap. In contrast, in the QAH phase, the edge state and
the zero modes are all stable.

\begin{figure}[tbph]
\includegraphics[width = 10.0cm]{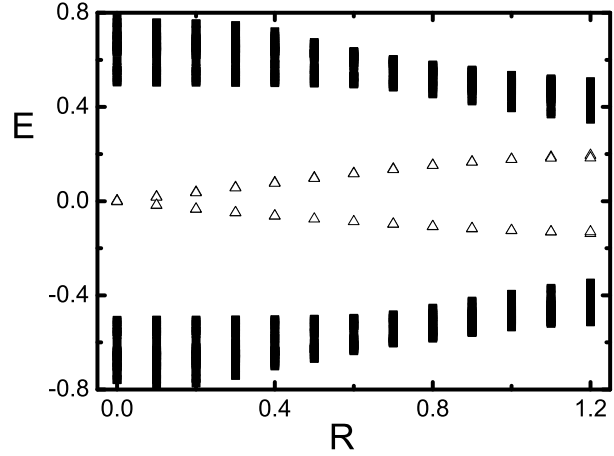}
\caption{The evolution of the energy spectrum of the system with a $\protect%
\pi $-flux at $D=0.5,$ $t=0.75$.}
\end{figure}

\begin{figure}[tbph]
\includegraphics[width = 12.0cm]{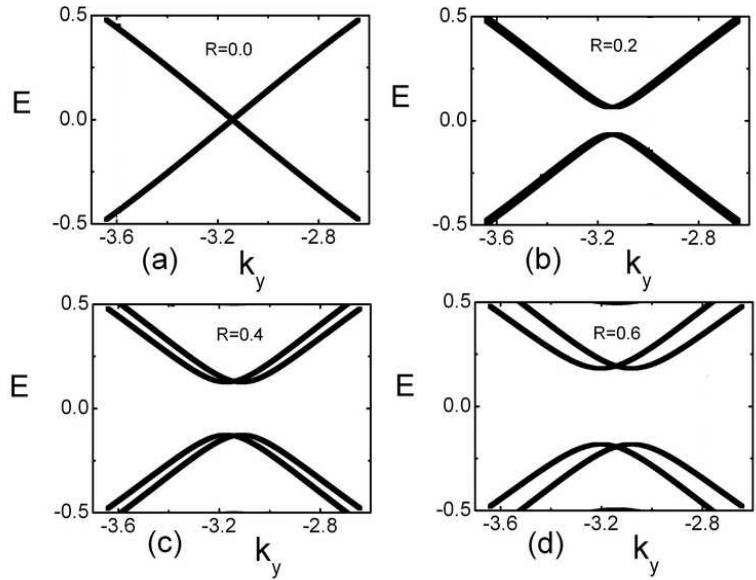}
\caption{The evolution of edge states in QSH state with the Rashba term $R$
at $D=0.5,$ $t=0.75$. a, b, c and d correspond to $R=0$, $R=0.2$, $R=0.4$
and $R=0.6$ respectively.}
\end{figure}

In this case (a system without T-symmetry and $S_{z}$-conservation), the
TKNN integer $Q$ can be regarded as an "order parameter" to characterize
different quantum states, $Q=-\frac{1}{2\pi }\int_{BZ}(\mathbf{\nabla }%
_{k}\times \mathbf{A})_{z}\,dk_{x}dk_{y}.$ Table.7 shows the relationship
between the TKNN integer $Q$ and the topological properties of the quantum
states,

\begin{equation*}
\begin{tabular}[t]{|c|c|c|}
\hline
$D\neq 0,$ $R\neq 0$ & QAH & \ NI\  \\ \hline
TKNN integer $Q$ & $1$ & $0$ \\ \hline
Edge state's number & $1$ & $0$ \\ \hline
Zero mode's number on $\pi $-flux & $1$ & $0$ \\ \hline
\end{tabular}%
.
\end{equation*}

Using similar approach in above sections, the universal critical behavior of
the TQPTs are studied. The TQPTs are\ also dominated by nodal Dirac
fermionic excitation at the point $(\pi ,$ $\pi )$. Since the TKNN integer $%
Q $ (the Chern number) will jump crossing the TQPTs, $\Delta Q=1,$ the mass
sign of one low energy fermionic excitation changes. Consequently, the TQPTs
are also third order.

\section{Conclusion}

In this paper, based on a two-dimensional lattice model, we study the TQPTs
between the QSH state, QAH state and normal band insulator and show their
physical properties, including the edge state, the quantized (spin) Hall
conductivity and the induced quantum number on a $\pi $-flux. There are
common features of the TQPTs for different cases : \emph{the existence of
nodal fermions at high symmetry points, the non-analytic third derivative of
ground state energy and the jumps of the topological "order parameters"}. In
particular, we find the \emph{symmetry protected nature} of the TQPTs which
is illustrated in Table.8 :%
\begin{equation*}
\begin{tabular}[t]{|c|c|c|}
\hline
T-symmetry & $S^{z}$-conservation & "Order parameter" \\ \hline
$\checkmark $ & $\times $ & $Z_{2}$ topological invariant $(-1)^{\Delta }$
\\ \hline
$\times $ & $\checkmark $ & Spin Chern number $Q_{s}$ \\ \hline
$\times $ & $\times $ & TKNN integer $Q$ \\ \hline
\end{tabular}%
.
\end{equation*}%
In Table.8 the symbol $\checkmark $ means the system is invariant of the
given symmetry and $\times $ means not. From Table.8, one can see that for a
system with $S^{z}$-conservation but without T-symmetry, the spin Chern
number $Q_{s}$ plays the role of "order parameter" to characterize the
topological insulators; for a system with T-symmetry but without $S^{z}$%
-conservation, the $Z_{2}$ topological invariant $(-1)^{\Delta }$ becomes
the "order parameter"; for the system without $S^{z}$-conservation and
T-symmetry, it is the TKNN integer $Q$ that becomes the "order parameter".

In the end, we give a comment on the relationship between symmetry and
topological invariants for the TQPTs. TQPTs are not be classified by
symmetries, instead, they may be characterized by some topological
invariants, such as the Chern number or $Z_{2}$ topological invariant.
However, the "\emph{symmetry}" of the systems still plays important role :
\emph{different topological quantum phase transitions are protected by
different (global) symmetries and then described by different topological
"order parameters"}.

\vskip 0.5cm

The authors thank Ying Ran and Xiao-Liang Qi for helpful discussions and
comments. The authors acknowledges that this research is supported by NCET,
NFSC Grant no. 10874017.


\begin{thebibliography}{99}
\bibitem{com} Since the universal properties of the TQPTs are beyond the
detail features of the lattice models. When one may begin with other lattice
models of topological insulators, they will find that the main results will
not change.
\end{thebibliography}
\end{document}